\documentclass[superscriptaddress,aps,apl,twocolumn,noeprint]{revtex4-1}
\usepackage{amsmath}
\usepackage{upgreek}
\usepackage{mathtools}
\usepackage{graphicx}
\usepackage{afterpage}
\usepackage{amssymb}
\usepackage{epstopdf}
\usepackage{hyperref}
\usepackage{tikz}
\usepackage{color}
\usepackage{bm}
\usepackage[caption=false,position=top,singlelinecheck=off,justification=raggedright]{subfig}

\graphicspath{{Images/}}

\newcommand*{\CC}{%
  \textsf{C\kern-1ex C}%
}

\DeclarePairedDelimiter\bra{\langle}{\rvert}  
\DeclarePairedDelimiter\ket{\lvert}{\rangle}       
\DeclarePairedDelimiterX\braket[2]{\langle}{\rangle}{#1 \delimsize\vert #2}      
\newcommand*{\be}{\begin{equation}}
\newcommand*{\ee}{\end{equation}}
\newcommand{\ba}{\[\begin{align}}
\newcommand{\ea}{\end{align}\]}       

\begin{document}

\title{Time-Evolving Weiss Fields in the Stochastic Approach to Quantum Spins}
\author{S. E. Begg}
\affiliation{Department of Physics, King's College London, Strand, London WC2R 2LS, United Kingdom}
\author{A. G.  Green}
\affiliation{London Centre for Nanotechnology, University College London,
Gordon St., London, WC1H 0AH, United Kingdom}
\author{M. J. Bhaseen}
\affiliation{Department of Physics, King's College London, Strand, London WC2R 2LS, United Kingdom}
\date{\today}
\begin{abstract}
We investigate non-equilibrium quantum spin systems via an exact
mapping to stochastic differential equations.  This description is
invariant under a shift in the mean of the Gaussian noise. We show
that one can extend the simulation time for real-time 
dynamics in one and two dimensions by a judicious choice of this
shift. This can be updated dynamically in order to reduce the impact
of stochastic fluctuations.  We discuss the connection to drift gauges
in the gauge-P literature.
\end{abstract}

\maketitle

\section{Introduction}
Quantum spin systems play a ubiquitous role in condensed-matter
physics, with a myriad of applications ranging from magnetic materials
to quantum computers.
Out of equilibrium, they exhibit a wealth of phenomena including anomalous
thermalization in low-dimensions \cite{Rigol2007,Rigol2008}
and dynamical quantum phase transitions
\cite{Heyl,Heyl2017}. In one-dimension (1D) they permit especially
strong links between theory and experiment, as exemplified by the
recent observation of dynamical quantum phase transitions using a 1D
chain of trapped ions \cite{Jurcevic2017}. They have also been
instrumental in the development of numerical algorithms, including
time-dependent Density Matrix Renormalization Group (tDMRG) and tensor
network approaches \cite{White2004,Vidal2004,Haegeman}. These methods
have enjoyed widespread applications in 1D, but they are much harder
to apply to non-equilibrium problems in higher dimensions. For state
of the art progress in this direction see for example
\cite{Paeckel2019,Czarnik2019,Hubig2020,Zaletel2020}.

Recently, an exact mapping between quantum spin dynamics and classical
stochastic differential equations (SDEs) has emerged, based upon the
Hubbard--Stratonovich decoupling of the exchange interactions \cite{Hogan2004,Galitski2011,Ringel2013,DeNicola2019,DeNicola2019long,Begg2019,DeNicola2019euclid}. This
stochastic approach allows for the numerical evaluation of
time-dependent quantum observables, in addition to analytical insights
obtained from the classical stochastic formulae \cite{DeNicola2019,DeNicola2019long,DeNicola2019euclid,Begg2019}. A notable feature is that it treats integrable and non-integrable
problems on a similar footing, including those in higher-dimensions. 
It also offers opportunities for developing links to a diverse body of phase space approaches
which have attracted attention in recent years \cite{Drummond1980,Deuar2002,Barry2008,Ng2011,Ng2013,Mandt2015,Wuster2017,Deuar2021,Steel1998,Polkovnikov2011,Schachenmayer2015,Schachenmayer2015b,Khasseh2020,Huber2020,Verstraelen2018,Verstraelen2020}.
In previous work \cite{Begg2019}, we
showed that the stochastic approach to quantum spins could be significantly improved by a two-patch
parameterization of the Bloch sphere, in conjunction with a
higher-order numerical integration scheme. We also highlighted the
link between the onset of stochastic fluctuations and the
non-Hermiticity of the effective stochastic Hamiltonian. 

\begin{figure}[t]
  \includegraphics[width=8.7cm]{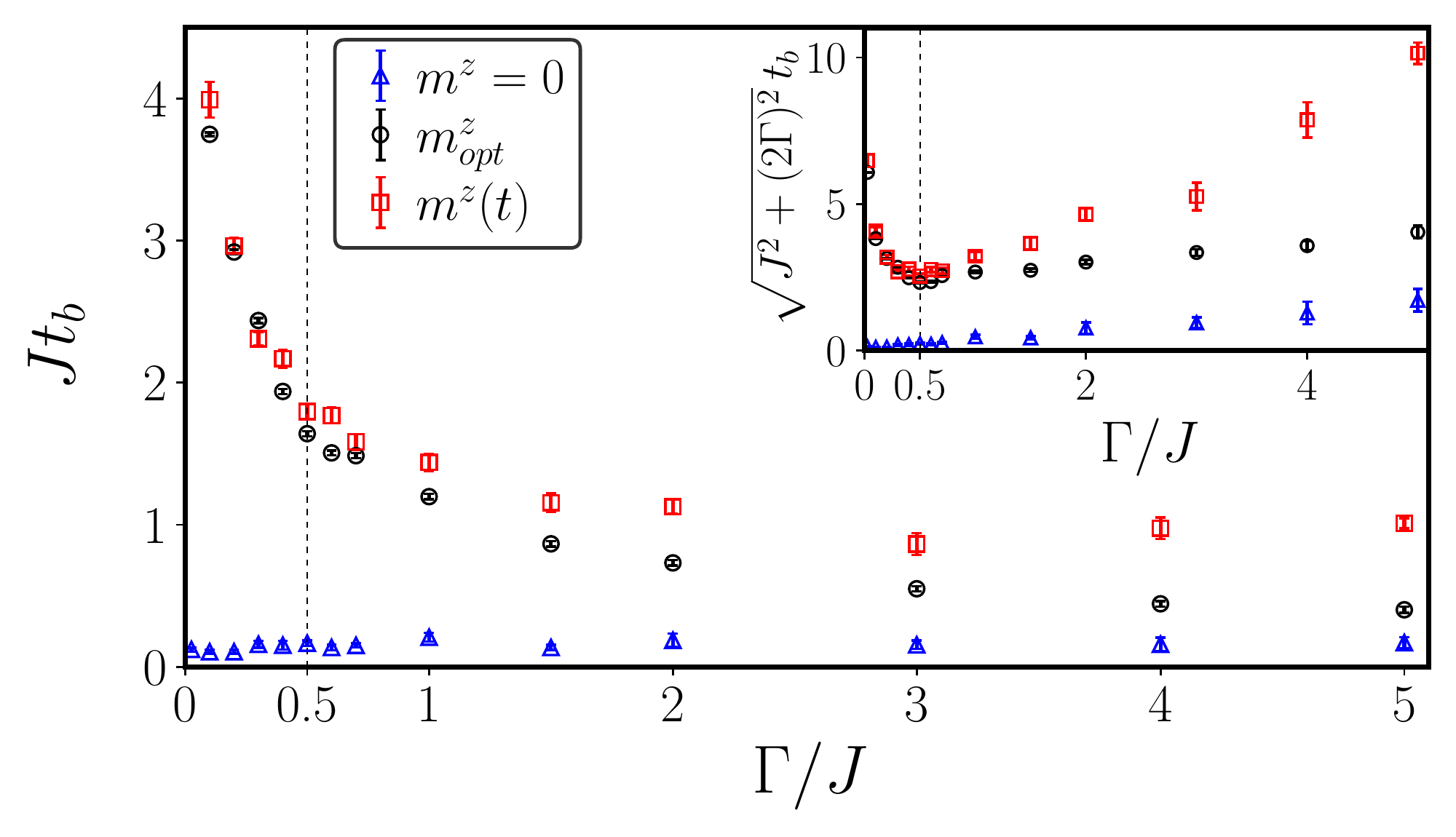}
\caption{Comparison of the breakdown time $t_b$, for simulations of
  the 1D quantum Ising model with $10$ spins, following a quantum
  quench from the fully-polarized initial state $\ket{\Downarrow}$, to
  different values of $\Gamma/J$. The results are obtained by
  numerical solution of the SDEs without a Weiss field ($m^z=0$), in
  the presence of an optimal static Weiss field ($m^z_{\rm opt}$), and
  with a time-evolving Weiss field $m^z(t)$. The use of a Weiss field
  leads to longer simulation times. Inset: rescaling $t_b$ by
  $\sqrt{J^2 + (2\Gamma)^2}$, which is proportional to the
  Hilbert--Schmidt norm of the Ising Hamiltonian $||\hat H_I||$,
  facilitates the comparison of $t_b$ for different values of
  $\Gamma/J$. With this rescaling, the smallest breakdown time, for
  both static and time-evolving Weiss fields, occurs at the
  critical point $\Gamma=J/2$.}
\label{fig:time_vs_phase}
\end{figure}

In this work, we show that the method for real-time dynamics can be further improved by the
use of a dynamical Weiss field to reduce the effects of
non-Hermiticity and stochastic fluctuations \cite{Begg2019}. In
essence, the Weiss field tracks the mean-field dynamics of the quantum
spin system, which facilitates more efficient sampling. Similar conclusions have been drawn in imaginary time using saddle-point techniques \cite{DeNicola2019euclid}. We demonstrate
these improvements by presenting results for the quantum Ising model
in both one and two dimensions, with up to 121 spins.  In the
Appendices, we discuss the link between the SDEs employed here,
and phase space methods using
  gauge-P density matrices \cite{Barry2008,Ng2013,Drummond1980,Deuar2001,Deuar2002}. We show that it is
possible to map between the two formalisms using a suitable choice of
drift gauge, previously considered for bosonic systems
\cite{Deuar2001,Deuar2002,Drummond2003,Drummond2004,Wuster2017}.  We conclude with directions for future
research.

\section{Stochastic Approach}
Recalling the principal steps of
Refs~\cite{Hogan2004,Ringel2013,DeNicola2019,DeNicola2019long,Begg2019},
the stochastic approach can be applied to a generic quadratic spin
Hamiltonian
\begin{equation} \hat{H} = -\frac{1}{2} \sum_{ijab}  J^{ab}_{ij} \hat{S}^a_i  \hat{S}^b_j - \sum_i h^a_i \hat{S}^a_i , \label{eq:heisenbergham}\end{equation}
where $J_{ij}^{ab}$ is the interaction between spins at
lattice sites $i,j$ and $h_i^a$ is an applied magnetic field.  The
spin operators, $\hat{S}^a_i$, obey the canonical commutation
relations $[\hat{S}^a_i,\hat{S}^b_j] = i \epsilon^{abc}\delta_{ij}
\hat{S}^c_j$, where $a,b ~ \epsilon ~ \{x,y,z\}$ label the spin
components, $\epsilon^{abc}$ is the antisymmetric symbol, and $\hbar =
1$.  The interactions in the corresponding time-evolution operator
$\hat{U}(t_f,t_i) = \mathbb{T}e^{-i \int_{t_i}^{t_f} \hat{H}(t) dt}$,
can be decoupled by performing a Hubbard--Stratonovich transformation
over auxiliary fields $\varphi_j^a$:
\begin{align}  \hat{U}(t_f,t_i)\! =\! \mathbb{T} \int \! \mathcal{D}\varphi  ~ e^{- S[\varphi] +  i\int\limits_{\mathclap{t_i}}^{\mathclap{t_f}}\! dt \sum_{ja} \Phi_j^a \hat{S}^a_{j}} ,\label{eq:HStransf} \end{align}
where ${\mathbb T}$ denotes time-ordering. Here,
$\mathcal{D}\varphi=\prod_{ja} {\mathcal D}\varphi_j^a$ and $\Phi_j^a =
\frac{1}{\sqrt{i}}\varphi_j^a + h_j^a\in {\mathbb C}$ plays the role
of an effective, complex magnetic field. The path integral weight 
\begin{align} S[\varphi] = \frac{1}{2}
  \int_{t_i}^{t_f} dt \sum_{ijab} \varphi^a_i~(J^{-1})^{ab}_{ij}
  \varphi^b_{j} \label{eq:whitenoisemeasure},\end{align} is referred
to as the {\em noise action} \cite{DeNicola2019long}, since it allows one to interpret the
fields $\varphi_j^a$ as Gaussian distributed random variables.
The problem therefore reduces to the dynamics of individual spins
coupled to noisy complex fields, where the decoupled spins evolve
under the stochastic Hamiltonian, $\hat{H}^s \equiv - \sum_{ja}
\Phi_j^a \hat{S}^a_j$. The spatial and
  temporal correlations between the spins are encoded in the
  correlations of the noise fields. By diagonalizing the noise action
\cite{Ringel2013,DeNicola2019,DeNicola2019long,Begg2019} one may
introduce new white noise variables $\phi_j^b$, \textit{via}
$\varphi^a_i \rightarrow \sum_{jb} O^{ab}_{ij} \phi_j^{b}$, where
$\bm{O}^T\bm{J}^{-1}\bm{O}= \bm{1}$; here we recast $O^{ab}_{ij}$ and
$J^{ab}_{ij}$ in terms of matrices $\bm{O}\equiv O_{(ai)(bj)}$ and
$\bm{J}\equiv J_{(ai)(bj)}$, where $(ai)$ is a two-component index.

The stochastic Hamiltonian gives rise to a stochastic evolution
operator $\hat{U}^s(t) = \mathbb{T} e^{-i \int_0^t \hat{H}^{s}(t')
  dt'}\equiv \prod_j \hat{U}_j^s(t)$, which factorizes into on-site
contributions. Using the Lie algebraic structure of $\hat{H}^s $ we
may parameterize $\hat{U}^{s}_j(t)$ \textit{via} a so-called
disentanglement transformation: $ \hat{U}^{s}_j(t) = e^{\xi^+_j(t)
  \hat{S}^+_j} e^{\xi^z_j(t) \hat{S}^z_j} e^{\xi^-_j(t) \hat{S}^-_j}$
\cite{Ringel2013}. The $\xi$-variables evolve according to stochastic
differential equations (SDEs):
\begin{subequations}
\label{eq:SDEs}
\begin{align}
 & -i \dot{\xi}^+_j = \Phi^+_j + \Phi^z_j \xi^+_j - \Phi^-_j \xi^{+^2}_j,        \label{eq:plus} 
\\ & -i \dot{\xi}^z_j = \Phi^z_j-2 \Phi^-_j \xi^+_j, \label{eq:zequat} \\ &                                                       
-i \dot{\xi}^-_j = \Phi^-_j e^{\xi^z_j},  \label{eq:mininit}  
\end{align}  
\end{subequations}
where $\Phi^{\pm}_j = \frac{1}{2} (\Phi^x_j \mp i \Phi^y_j)$. The latter can be written in terms of the white noise variables as  $\Phi^a_j = \frac{1}{\sqrt{i}}\sum_{jb} O^{ab}_{ij} \phi_j^{b} + h_j^a, $ where \cite{Ringel2013,DeNicola2019,DeNicola2019long,Begg2019} 
\begin{align}
\langle \phi^a_i(t) \phi^b_j(t') \rangle = \delta_{ab}\delta_{ij}\delta(t-t'), ~~~ \langle \phi^a_i(t) \rangle =0.
\end{align} 
To calculate quantum observables, $\langle \hat{{\mathcal O}}(t) \rangle =
\bra{\psi(0)} \hat{U}^{\dagger} \hat{{\mathcal O}} \hat{U} \ket{\psi(0)}$, both
the forwards and backwards time-evolution operators must be
independently decoupled \cite{DeNicola2019}.
Observables thereby reduce to averages of functions of the associated
decoupling fields, $\xi$ and $\tilde{\xi}$ \cite{DeNicola2019}. To
solve the SDEs (\ref{eq:plus}) and (\ref{eq:zequat}), we use the Heun
predictor-corrector integration scheme in the Stratonovich formalism
\cite{Ruemelin1982,Kloden1992}, with a time-step $dt = 0.01$, unless
stated otherwise. We also remove coordinate singularities \textit{via}
the two-patch approach given in \cite{Begg2019}.

\section{Effective Weiss Field}
A key feature of the representation (\ref{eq:HStransf}) is that it is
invariant under shifts of the Hubbard--Stratonovich fields
  $\varphi(t) \rightarrow \varphi(t) + \Delta \varphi(t)$, since the
  fields correspond to dummy integration variables in the path
  integral. This leaves the time-evolution operator unchanged, which was recently used to develop an importance sampling approach in imaginary time 
\cite{DeNicola2019euclid,DeNicolaThesis}. In this work, we show that a judicious
choice of $\Delta\varphi(t)$ can significantly improve numerical
simulations of real-time dynamics over a broad range of parameters.
To gain some intuition for this, we note that under this
transformation, the effective magnetic field transforms as $\Phi_i^a
\rightarrow \frac{1}{\sqrt{i}}\sum_{jb}O_{ij}^{ab}( \phi_j^b + \Delta
\phi_j^b) + h_i^a$.  Denoting $\Delta \phi_j^b = \sqrt{i}
\sum_{kc}m_k^c O^{cb}_{kj} $, this can be rewritten as $\Phi_i^a =
\frac{1}{\sqrt{i}}\sum_{jb}O_{ij}^{ab} \phi_j^b + h_i^a + \sum_{jb}
J_{ij}^{ab}m_j^b$. At this stage the parameter $m_j^b$ is completely
arbitrary.  However, as we will expand upon in Sections \ref{sec:static_weiss} and
  \ref{sec:dynamical weiss}, the contribution $\sum_{jb}
J_{ij}^{ab}m_j^b$ can be interpreted as an effective Weiss field due
to the neighboring spins. For example, in the special case of
isotropic nearest neighbor interactions, this
reduces to $Z J^{ab} m^b$, where $m^b=m_j^b$ and $Z$ is the
coordination number. This mirrors the mean field contribution of
neighboring spins to the local Weiss field, where $m_j^b$ is the
component of the magnetization in the direction specified by $b$. More
generally, we may choose the parameter $m_j^b(t)$ to be
time-dependent, in accordance with the dynamics of the neighboring
spins. The shift of the fields $\varphi$ also induces a transformation
of the probability measure \textit{via} the noise action
(\ref{eq:whitenoisemeasure}) \cite{DeNicola2019euclid,DeNicolaThesis}:
 \begin{align} S[\phi] \rightarrow S[\phi,m] = S[\phi]  + \Delta S[\phi,m],\end{align}
where 
\begin{align}\Delta S =   &\frac{1}{2}  \int_{t_i}^{t_f}  dt \Big(2\sqrt{i}\sum_{ijab}  m^{a}_{i}O_{ij}^{ab}\phi^{b}_{j}  + i \sum_{ijab}  J_{ij}^{ab}m^{a}_{i}m^{b}_{j}\Big) \label{eq:transformedmeasure},\end{align}
and $S[\phi]= \frac{1}{2}  \int_{t_i}^{t_f}  dt  \sum_{ia}\left(\phi^{a}_{i}\right)^2$ is the diagonal form of the noise action.
This re-weights the stochastic trajectories by terms involving the
dynamical Weiss field $m_i^a(t)$. 

In Fig.~\ref{fig:time_vs_phase} we highlight the improvements obtained
by the use of a Weiss field. The figure shows the breakdown time of
numerical simulations, $t_b$, following a quantum quench in the 1D
quantum Ising model
\begin{equation} 
\hat{H}_I =   - \frac{1}{2} \sum_{\langle ij \rangle}J_{ij}  \hat{S}^z_i \hat{S} ^z_{j}  - \Gamma \sum_{j=1}^{N}  \hat{S}^x_j, \label{eq:TFIM} 
\end{equation}
with $N=10$ spins and nearest neighbor interactions $J_{ij} = J$, from the fully-polarized state
$\ket{\Downarrow}\equiv \prod_j\ket{\downarrow}_j$ to different
values of $\Gamma/J$. The relevant SDEs are  
\begin{subequations}
\label{eq:isingSDEs}
\begin{align}
 & -i \dot{\xi}^+_j = \frac{\Gamma}{2} + \Big(\frac{1}{\sqrt{i}}\sum_{k} O^{zz}_{jk}\phi_k^{z} + \sum_kJ_{jk}m_k^z\Big)  \xi^+_j - \frac{\Gamma}{2} \xi^{+^2}_j,        \label{eq:isingplus} 
\\ & -i \dot{\xi}^z_j = \frac{1}{\sqrt{i}}\sum_kO^{zz}_{jk} \phi_k^{z}+ \sum_kJ_{jk}m_k^z  - \Gamma \xi^+_j, \label{eq:isingzequat} \\ &                                                       
-i \dot{\xi}^-_j = \frac{\Gamma}{2} e^{\xi^z_j}.  \label{eq:isingmininit} 
\end{align}  
\end{subequations}
In practice, the variable $\xi^-_j$ can be neglected, since it
  drops out of observables involving the initial spin down state at site $j$
  \cite{Begg2019}.
The data in
Fig.~\ref{fig:time_vs_phase} correspond to (i) the SDEs without
a Weiss field ($m_j^z = 0$); (ii) an optimal choice of spatially uniform static
Weiss field, as discussed in Section \ref{sec:static_weiss}); and
(iii) a spatially uniform time-evolving Weiss field, $m^z(t)$,
which is determined self-consistently in Section \ref{sec:dynamical
  weiss}. The key point, is that the use of a Weiss field leads to
longer breakdown times, over a broad range of parameters.
In the remainder of this work, we will consider each of these cases in
turn. In Section \ref{sec:static_weiss} we consider the case where
$m_j^b$ is spatially homogeneous and static, and investigate its
impact upon the dynamics of quantum expectation values. In Section
\ref{sec:dynamical weiss} we consider time-dependent extensions \textit{via} a
self-consistent choice of $m_j^b(t)$. In the Appendices, we demonstrate
that the generalized SDEs, including a Weiss field, can be obtained
within the gauge-P approach for a particular choice of drift gauge.

\begin{figure}[t]
\subfloat{
\includegraphics[width =8.7cm]{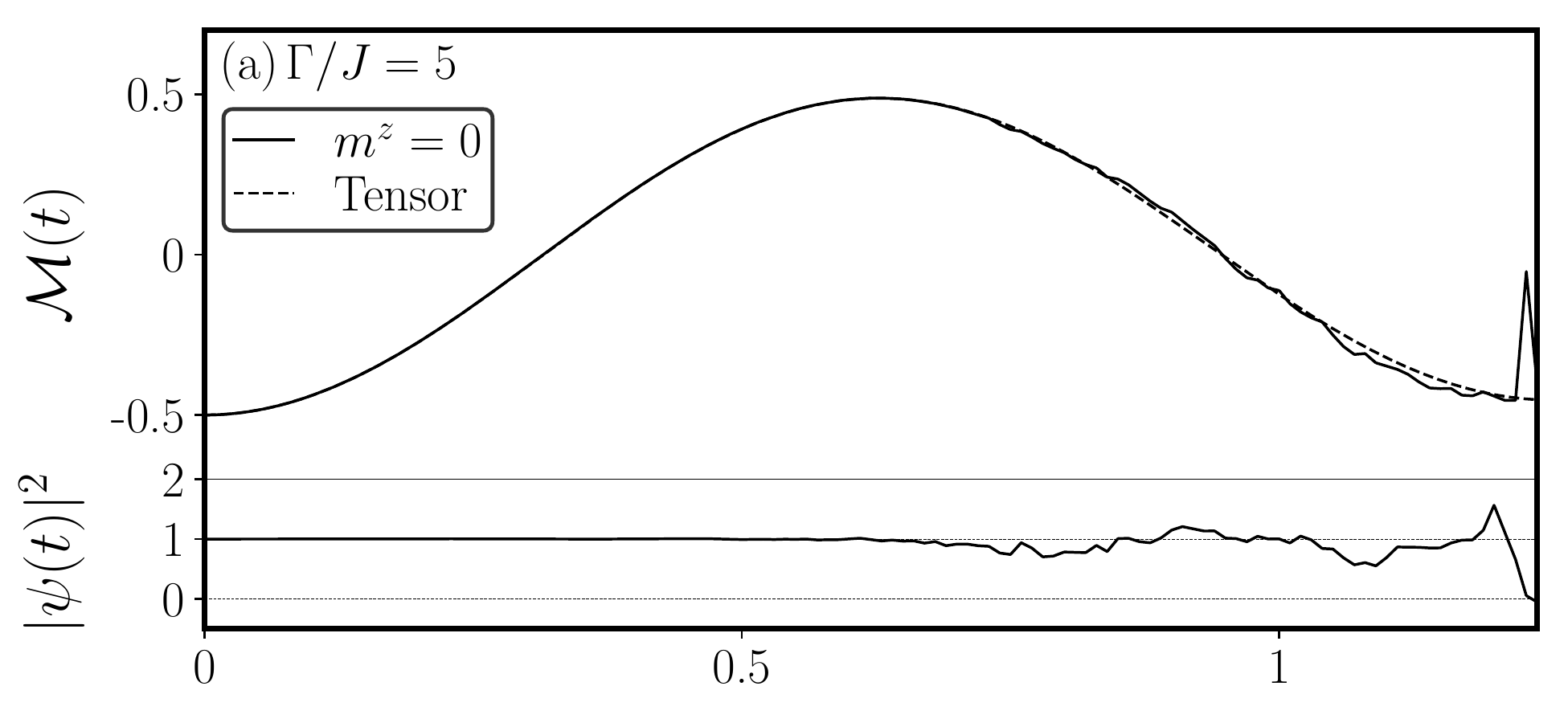}}

\vspace{-0.3cm}
\subfloat{
\includegraphics[width =8.7cm]{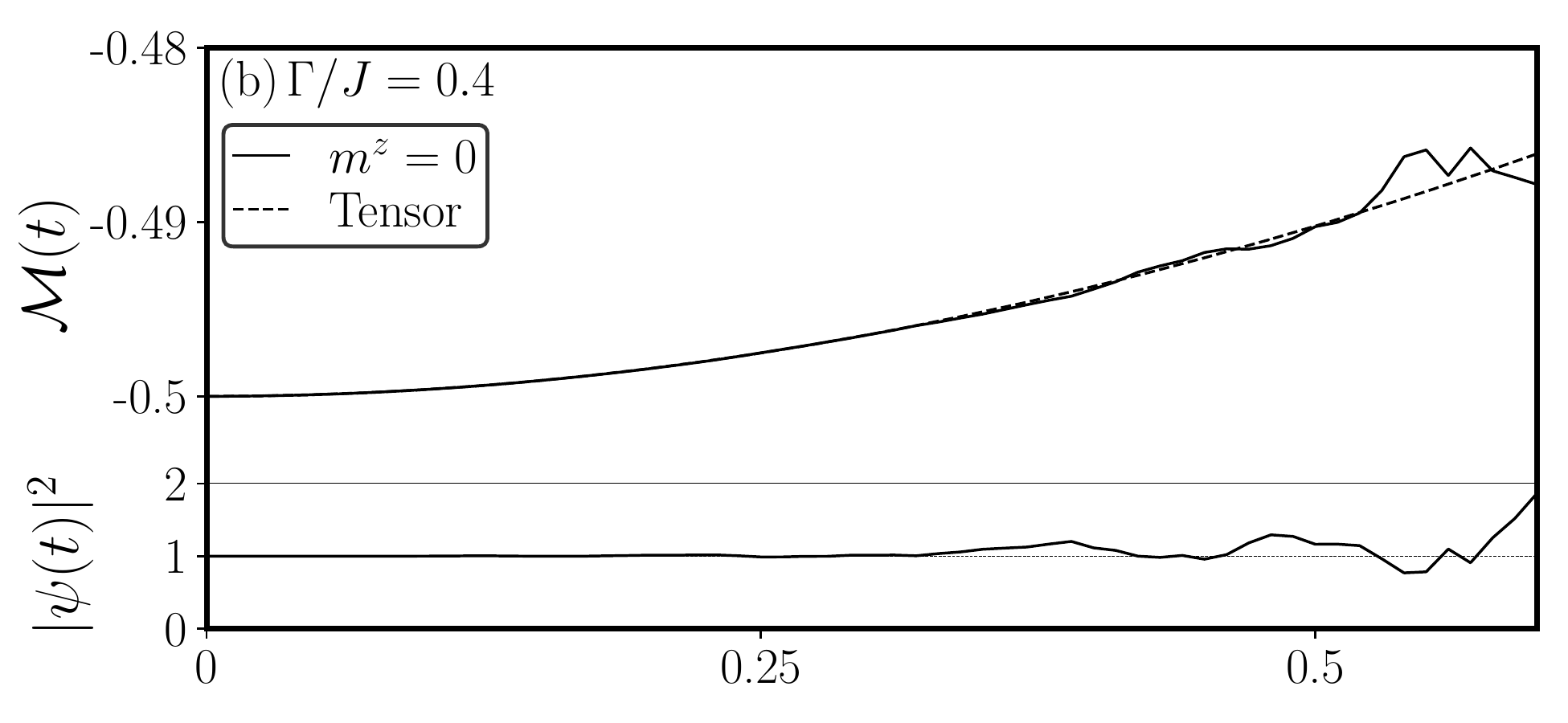}}
\centering

\vspace{-0.3cm}
\subfloat{
\includegraphics[width =8.74cm]{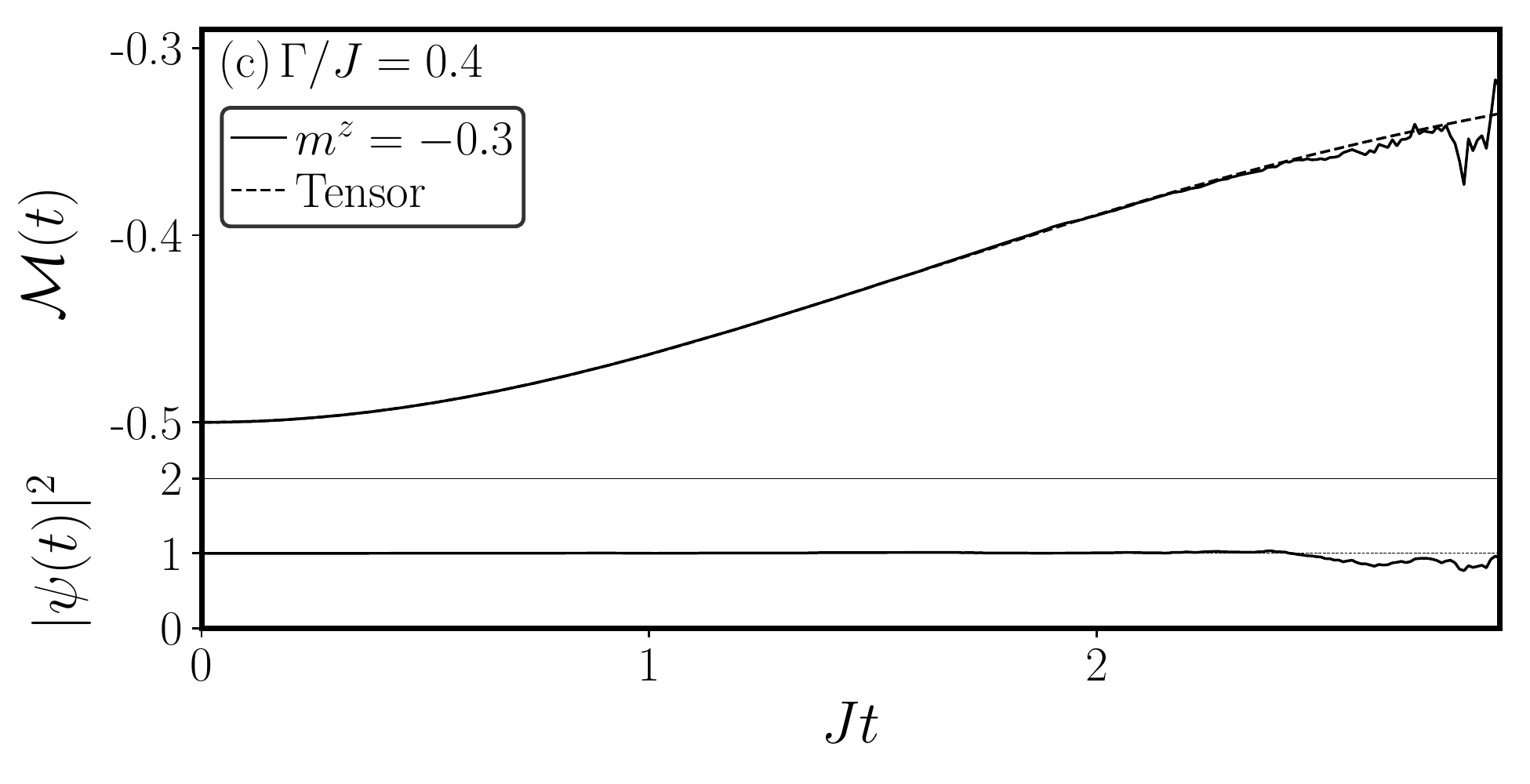}}
\caption{Time-evolution of the magnetization $\mathcal{M}(t)$
  following a quantum quench from the fully polarized initial state
  $\ket{\Downarrow}$ in the 1D quantum Ising model with $25$
  sites. (a) Quench to $\Gamma/J = 5$ using the SDEs in the absence of
  a Weiss field with $m^z=0$ (solid) and using tensor network Matrix Product Operator (MPO) methods
  (dashed). (b) Quench to $\Gamma/J = 0.4$ with $m^z = 0$ (solid) and
  using tensor networks (dashed). The accessible timescale for the
  stochastic approach is reduced in comparison to (a). This can be
  extended by using a well-chosen static Weiss field. (c) Quench to
  $\Gamma/J = 0.4$ with $m^z = -0.3$ (solid) showing improved
  simulation times. In all the cases we average over $\mathcal{N} =
  10^6$ trajectories. The norm of the quantum state is also shown to
  indicate the reliability of the simulations, and the eventual
  breakdown time.}
\label{fig:performance}
\end{figure}

\section{Static Weiss Field} \label{sec:static_weiss}
In this section we explore the improvements in numerical simulations
obtained through the use of a static Weiss field. We consider quantum
quenches in the 1D quantum Ising model (\ref{eq:TFIM}), with periodic
boundary conditions and $J=1$. We start in the fully-polarized initial
state $\ket{\Downarrow} = \prod_{i=1}^N \ket{\downarrow}$, and quench
to different values of $\Gamma/J$. The expectation value of the spin
operator $\hat{\mathbf{S}}_j$ has an intuitive representation in the
stochastic approach \cite{Begg2019}: 
\begin{align} \langle
  \hat{\mathbf{S}}_j(t) \rangle = \Big\langle {\mathcal W}\prod_i
  |\psi_i^s(t)|^2
  \mathbf{n}_j(t)\Big\rangle_{\phi,\tilde{\phi}}, \label{eq:singlespin}\end{align}
where $\langle ... \rangle_{\phi,\tilde\phi}$ denotes averaging
  over the Gaussian white noise variables. The weight ${\mathcal
    W}=e^{-\Delta S[\phi,m] - \Delta S^*[\tilde{\phi},m]}$ is
  discretized in time and it weights the stochastic trajectories {\em via}
  the Weiss-field. This mirrors the re-weighting of trajectories performed in imaginary time \cite{DeNicola2019euclid}. The vector $\mathbf{n}_j(t)$ corresponds to the
position of a spin on the Bloch sphere, expressed in terms of
projective coordinates \cite{Begg2019}:
\begin{align}\mathbf{n}_j(t) =
  \frac{1}{2}\Bigg(\frac{2\text{Re}(\xi^+_j(t))}{1+ |\xi^+_j(t)|^2}
  ,\frac{-2\text{Im}(\xi^+_j(t))}{1+ |\xi^+_j(t)|^2} ,\frac{-1 +
    |\xi^+_j(t)|^2}{1+ |\xi^+_j(t)|^2}
  \Bigg). \label{eq:euclid} \end{align} The factor of
$|\psi_i^s(t)|^2$ corresponds to the norm of the stochastic state
$\ket{\psi_i^s(t)} =\hat{U}_i^s(t) \ket{\psi(0)}$, and is given by
\begin{align}
  |\psi_i^s(t)|^2 = e^{-\text{Re}(\xi^{z}_i(t))}(1 +
  |\xi^+_i(t)|^2). \label{eq:normstoch}
\end{align}  In writing (\ref{eq:euclid}) and (\ref{eq:normstoch}), it is implicit that the conjugate variable $\xi^{a*}_j$ is
independent of $\xi^{a}_j$; we denote this {\em via} the replacement
$ \xi^{a*}_j \rightarrow  \tilde{\xi}^{a*}_j$. 
  Although (\ref{eq:singlespin}) is formally exact, the norm of the
  quantum state is not preserved in numerical simulations with a
  finite number, $\mathcal{N}$, of stochastic samples
  \cite{Begg2019}. As such, we further rescale by the quantum state
  norm \cite{Begg2019} \begin{align} |\psi(t)|^2 = \Big\langle
    {\mathcal W}\prod_i |\psi_i^s(t)|^2
    \Big\rangle_{\phi,\tilde{\phi}}\,. \label{eq:onsitenorm}\end{align}

\begin{figure}[t]

\subfloat{\includegraphics[width = 8.7cm]{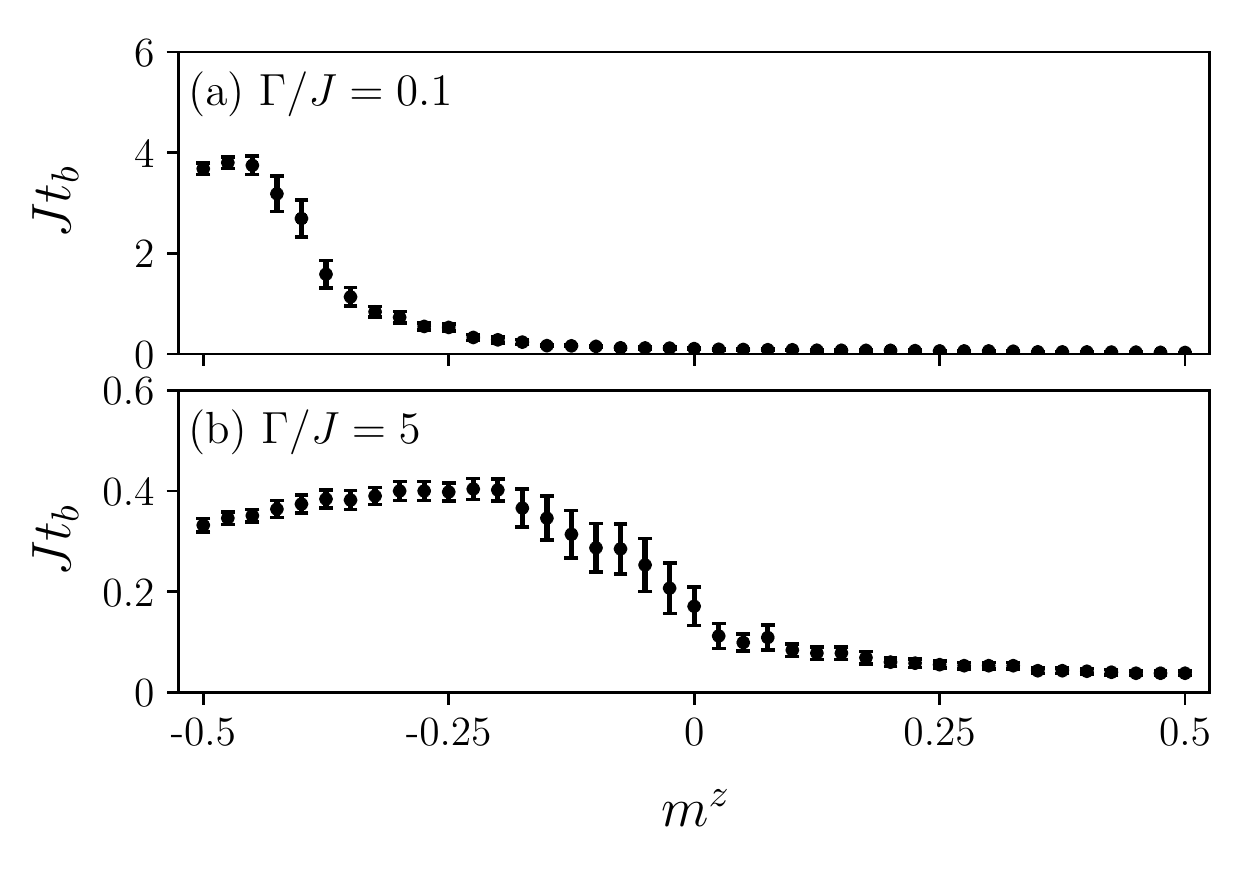}}

\vspace{-0.6cm}
\subfloat{\includegraphics[width =8.4cm]{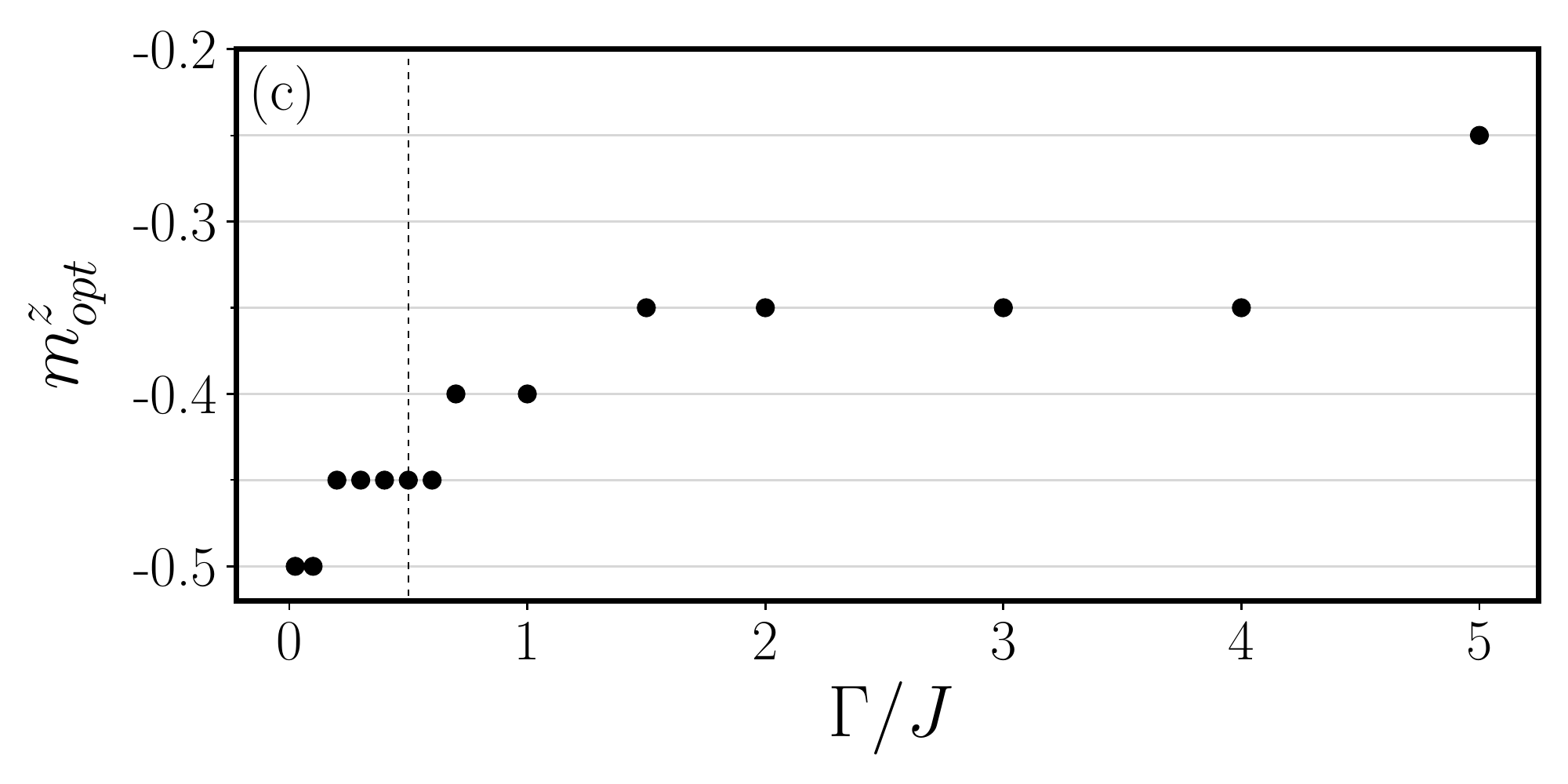}}
\caption{Breakdown time, $t_b$, versus the static Weiss field, $m^z$,
  for quenches in the 1D quantum Ising model with $10$ spins, from the
  fully-polarized state $\ket{\Downarrow}$ to (a) $\Gamma/J =0.1$ and
  (b) $\Gamma/J=5$. The optimal Weiss field, corresponding to the
  longest simulation time, is given by $m^z_{\rm opt} \approx - 0.45$
  in (a) and $m^z_{\rm opt} \approx -0.2$ in (b). (c) Variation of
  $m^z_{\rm opt}$ as a function of the post-quench value of $\Gamma/J $.
  The optimal value is selected from the range $-0.5 \leq m_{\rm
    opt}^z \leq 0$, which is discretized in steps of 0.05. The
  breakdown times are obtained as the average of the breakdown time
  from 10 batches of $\mathcal{N} = 10,000$ runs. The error bars
  in (a) and (b) correspond to the standard error of
  these batches.}
\label{fig:timescales_vs_m}
\end{figure}

In Fig.~\ref{fig:performance}(a) we show the time-dependence of the
magnetization, $\mathcal{M}(t) = \frac{1}{N} \sum_{j=1}^N \langle
\hat{S}_j^z \rangle$ with $N=25$, following a quantum quench from the
initial state $\ket{\Downarrow}$ to the paramagnetic phase with
$\Gamma=5J$. The results are obtained in the absence of a Weiss field
($m_j^z=0$) and are in excellent agreement with those obtained \textit{via} the tensor network 
Matrix Product Operator technique MPO $W^{I}$ \cite{Zaletel2015} for the
time-interval displayed. For comparison, we also show the norm of the
time-evolving quantum state as calculated \textit{via}
(\ref{eq:onsitenorm}). It is readily seen that departures from
coincidence occur when the norm deviates from unity \cite{Begg2019}. Throughout this
work, we define the breakdown time $t_b$ of our numerical simulations,
as the time at which this deviation reaches $1\%$.
In Fig.~\ref{fig:performance}(b), we show results for a quench to
$\Gamma = 0.4J$, within the ferromagnetic phase. The results are obtained in the absence of a Weiss field ($m_j^z=0$) and break down at
an earlier time than those in panel (a).
For comparison, in Fig.~\ref{fig:performance}(c) we show results for
the same quench as in panel (b), but in the presence of a
static Weiss field $m_j^z = -0.3$; as we will discuss below, this turns
out to be a near-optimal choice of the static Weiss field, for this
particular quench. Since the model (\ref{eq:TFIM}) only
contains z-interactions, we consider Weiss fields in the
z-direction only. It is evident that the simulation time is
extended, beyond that in panels (a) and (b).

In order to gain some insight into the variation of the breakdown time
$t_b$ with $m^z=m_j^z$, we consider quenches to different points in
the phase diagram as a function of $m^z$. To aid the comparison, we
fix the number of stochastic samples to $\mathcal{N} = 10,000$. In
Fig.~\ref{fig:timescales_vs_m}(a) we plot the dimensionless breakdown
time $Jt_b$, versus $m^z$ for a quench to $\Gamma = 0.1J$ within the
ferromagnetic phase. It can be seen that the best choices for the
static Weiss field lie in the range $-0.5 \lesssim m^z \lesssim
-0.35$. In Fig. \ref{fig:timescales_vs_m}(b) we do the same analysis
for $\Gamma = 5J$. It can be seen that this larger value of $\Gamma$
reduces the magnitude of the optimal choice for $m^z$. In
Fig.~\ref{fig:timescales_vs_m}(c) we show the variation of the optimal
Weiss field, $m^z_{\rm opt}$ for quenches to different points in the
phase diagram. It can be seen that the optimal choice of $m^z$
interpolates between $m^z=-1/2$ and $m^z=0$ as one passes from the
ferromagnetic region ($\Gamma <J/2$) to the paramagnetic region
($\Gamma>J/2$).

In Fig.~\ref{fig:time_vs_phase} we show the breakdown time
corresponding to the optimal static Weiss field. It can be seen that
the use of a Weiss field leads to a significant improvement in the
simulation time throughout the phase diagram. The inset shows the same
data rescaled by $\sqrt{J^2 + (2\Gamma)^2}$, which is proportional to 
the Hilbert-Schmidt norm of the Ising Hamiltonian $||\hat{H}_I||_2=\sqrt{\text{Tr}(\hat H_I^2)}$ \cite{Horn2012}. This facilitates the comparison of the timescales
for different quantum quenches. It can be seen that the shortest rescaled
simulation times occur for quenches close to the quantum critical point
at $\Gamma=J/2$, as one would na\"ively expect due to enhanced fluctuations.

\begin{figure}[t]
\centering
\includegraphics[width = 8.7cm]{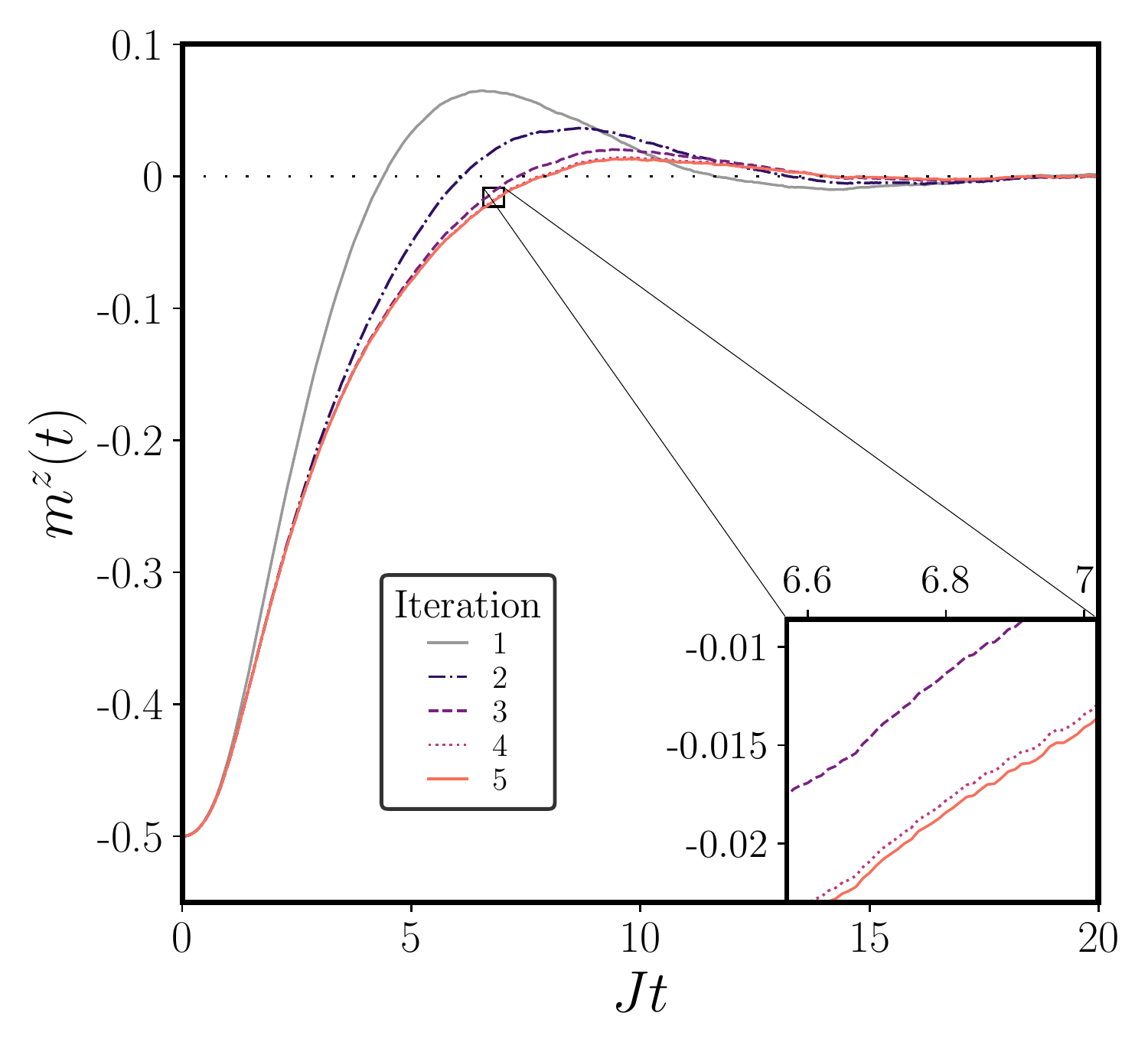}
\caption{Determination of the time-evolving Weiss field $m^z(t)$
  following a quantum quench in the 1D quantum Ising model with $10$
  sites, from the fully-polarized state $\ket{\Downarrow}$ to $\Gamma
  / J = 0.4$. The results correspond to five iterations of the
  procedure discussed in the main text, where each iteration
  corresponds to $\mathcal{N} = 10,000$ stochastic trajectories. As
  shown in the inset, the results converge to a fixed time-dependent
  profile for $m^z(t)$.}
\label{fig:weiss_iterations}
\end{figure}

\section{Time-Evolving Weiss Field} \label{sec:dynamical weiss}

In this section we examine the possibility of choosing the value of
$m_j^z$ as a function of time.  A natural choice is evident if we
write the stochastic Hamiltonian for the quantum Ising model in the
form
\begin{align} \hat{H}^{s}(t) = &- \sum_{i}  \Gamma \hat{S}^x_{i} - \sum_{ij} J_{ij} m_j^z(t) \hat{S}^z_{i} \nonumber \\ & \hspace*{0.2cm} - \frac{1}{\sqrt{i}} \sum_i  \varphi_i^z   \big(\hat{S}^z_i - m_i^z(t) \hat{\mathbb{I}} \big), \label{eq:hstochwithgauge} \end{align} where $\hat{\mathbb{I}}$ is the identity operator
and $\varphi_i^z$ is the original decoupling field with probability
measure (\ref{eq:whitenoisemeasure}).
The additional terms that would arise \textit{via}
(\ref{eq:transformedmeasure}) have been absorbed into
$\hat{H}^{s}(t)$; the ${\mathcal O}(m^2)$ terms can be neglected since
they result in a deterministic phase for $\ket{\psi^s(t)}$ which is identical for all trajectories. Choosing $m^z_j(t)$ to be the
instantaneous average of $\langle \hat S_j^z \rangle$ allows one to reduce the effects of
non-Hermiticity arising from (\ref{eq:hstochwithgauge}):
\begin{align} m_j^z(t) =  \Bigg\langle
    \frac{\bra{\psi^s(t)}\hat{S}_j^z\ket{\psi^s(t)}}{|\psi^s(t)|^2}\Bigg\rangle_{\phi}\label{eq:meanfieldit}, \end{align}
where the average is over the noise variables associated with the
forwards time-evolution; the Weiss field for the backwards evolution
takes the same value. Enforcing the Bloch-sphere normalization
explicitly in (\ref{eq:meanfieldit}) results in contributions to the
average that are comparable in size. The result therefore converges
with far fewer samples than are needed for quantum observables such as
(\ref{eq:singlespin}).  The choice (\ref{eq:meanfieldit}) also
generates the physically transparent mean-field term $\sum_{ij} J_{ij}
m^z_i \hat{S}^z_{j}$ in the stochastic Hamiltonian
(\ref{eq:hstochwithgauge}). 
This is analogous to the optimal shift for imaginary time evolution,  corresponding to a mean-field saddle-point \cite{DeNicola2019euclid}.
Since $\ket{\psi^s(t)}$ is itself a
function of $m_j^z(t)$, the Weiss field should be determined
iteratively. To do this, we first set $m_j^z(t)=0$ and simulate
trajectories to yield (\ref{eq:meanfieldit}). This is then used as
$m_j^z(t)$ for the next simulation. We proceed in this iterative
fashion until $m_j^z(t)$ converges to a fixed time-evolution.  For
translationally invariant states, we may consider a single Weiss field
$m^z(t)= \frac{1}{N} \sum_{j=1}^N m_j^z(t)$ applied to all the
sites. As we discuss in Appendix \ref{sec:weiss_est}, one can estimate
this field from a small sub-system that captures the local
interactions.

In Fig. \ref{fig:weiss_iterations} we plot $m^z(t)$ as a function of
time for simulations of the 1D quantum Ising model with $N=10$
spins. We consider a quantum quench from the fully-polarized state
$\ket{\Downarrow}$ to $\Gamma = 0.4J$, showing the results from each
iteration. After four iterations of ${\mathcal N}=10,000$ samples the
data converge to a fixed-point value of $m^z(t)$, to a high level of
accuracy. As shown in Fig. \ref{fig:time_vs_phase}, the time-evolving
Weiss field performs at least as well as the optimal static
choice. For small $\Gamma / J$, a key advantage of the time-dependent
procedure is that one does not have to survey different static Weiss
fields. For $\Gamma \gg J$ the performance of $m^z(t)$ is superior to
$m^z_{\rm opt}$, as it self-consistently tracks the mean-field
dynamics. In comparison, the optimal static Weiss field, $m^z_{\rm opt} =
0$, captures only the time-average of the time-evolving mean-field.

\begin{figure}[t]
\subfloat{\centering
\includegraphics[width =8.8cm]{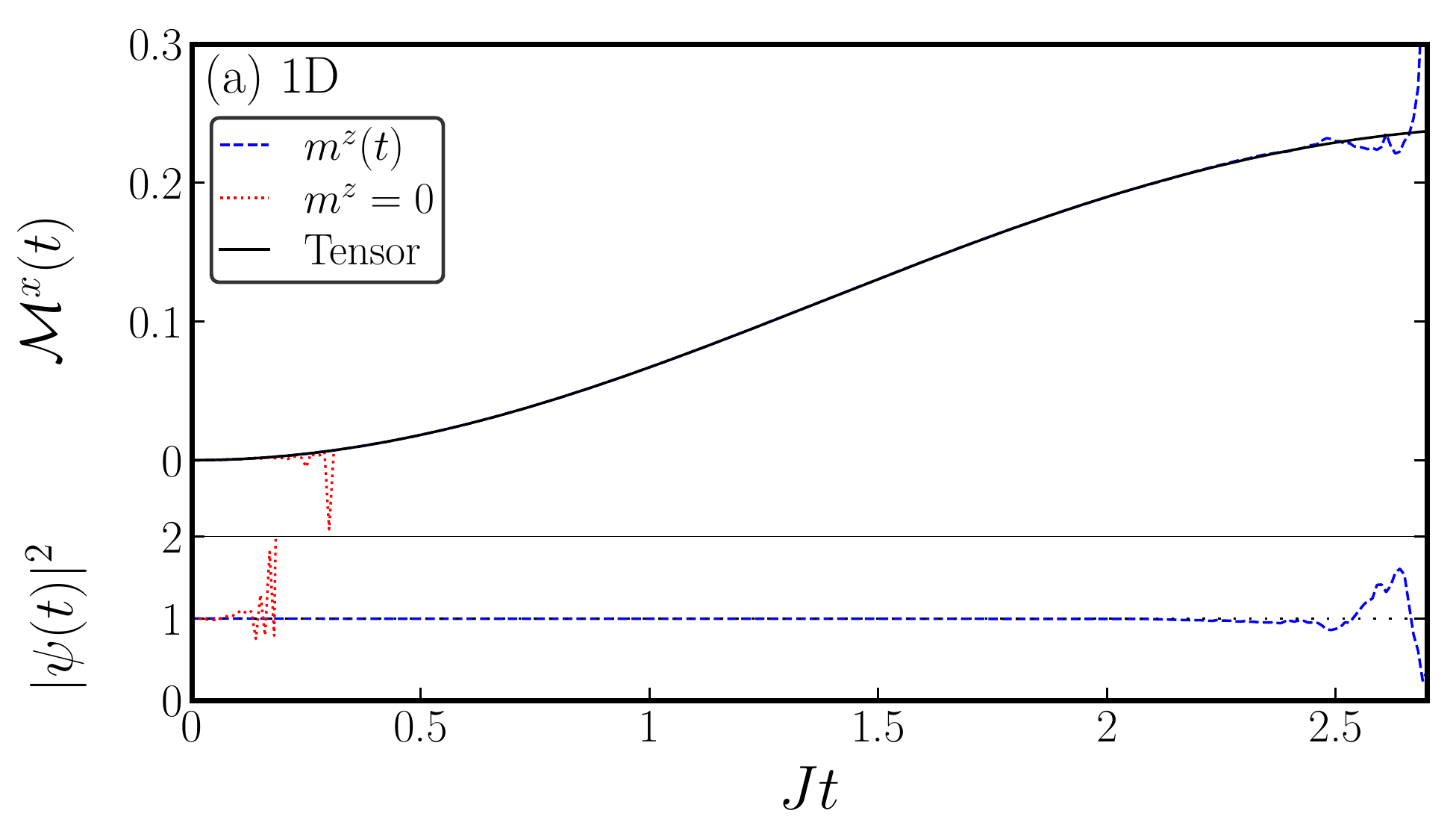}}

\vspace{-0.5cm}
\subfloat{\centering
\includegraphics[width =8.8cm]{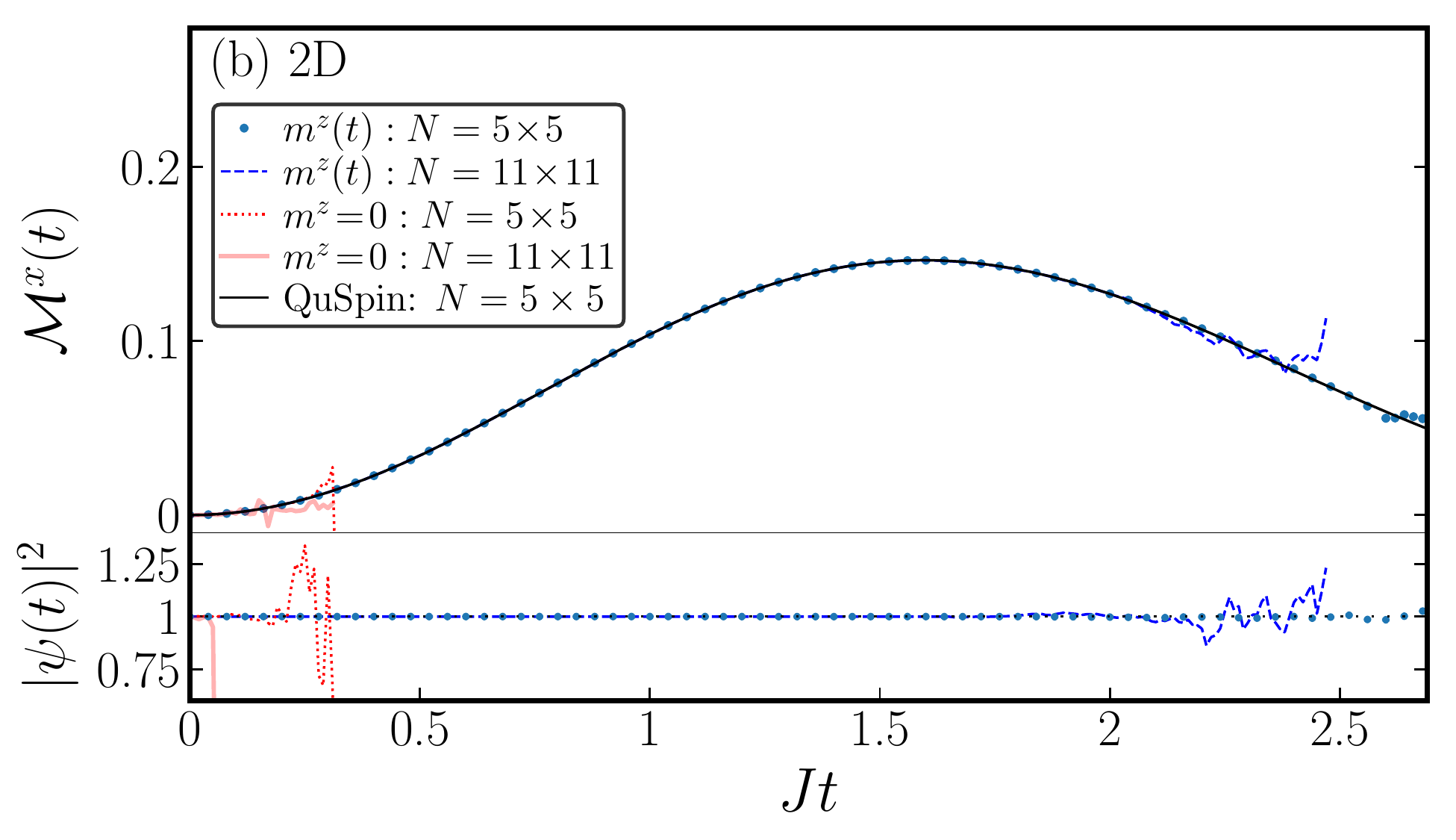}}
\caption{Dynamics of the transverse magnetization $\mathcal{M}^x(t)$
  following a quantum quench in the (a) 1D and (b) 2D quantum Ising
  model, from the the fully-polarized initial state $\ket{\Downarrow}$
  to $\Gamma/J = 0.3$. (a) 1D results for a $101$ site system obtained
  from the SDEs using a time-dependent Weiss field $m^z(t)$ and
  $\mathcal{N} = 5 \times 10^5$ trajectories (dashed). The results are
  in agreement with those obtained \textit{via} tensor networks
  (solid) until the breakdown time. It can be seen that the results
  for $m^z=0$ (dotted) break down earlier. (b) 2D results for a $5
  \times 5$ lattice with $\mathcal{N} = 10^6$ (dots) and an $11 \times
  11$ lattice with $\mathcal{N} = 5 \times 10^5$ (dashed). The former
  are in agreement with the results obtained \textit{via} QuSpin's ODE
  solver \cite{Weinberg2019} (solid). In the absence of a comparison
  to other techniques, the $11 \times 11$ results are seen to be in
  good agreement with the $5 \times 5$ results until the breakdown
  time; beyond this, strong fluctuations occur in the $11 \times 11$
  case. Once again, the $m^z = 0$ results (dotted and light solid)
  break down earlier.  In all cases, the results are plotted until
  fluctuations cause departures from the true dynamics. 
  In each panel, the time-dependent Weiss
field $m^z(t)$ is obtained by four iterations of the procedure
discussed in the main text using $\mathcal{N} = 10^3$ stochastic
samples.}
\label{fig:2D}
\end{figure}

\section{Implementation}
Having established a protocol for determining the time-evolving Weiss
field $m^z(t)$, we now explore its effectiveness in numerical
simulations. We focus on moderately large system sizes in both one and
two dimensions.  Throughout this section, the Weiss field is
determined by four iterations of the self-consistent approach with a
relatively small number of ${\mathcal N}=1000$ samples. In
Fig. \ref{fig:2D}(a) we show results for the transverse magnetization,
$\mathcal{M}^x(t) = \frac{1}{N}\sum_i \langle \hat{S}_i^x \rangle$, following a quantum quench from $\ket{\Downarrow}$
to $\Gamma = 0.3J$, in the 1D quantum Ising model with $N=101$
spins. The results are in very good agreement with tensor network
methods until $Jt_b = 2.11$; this is a significant improvement
over the $m^z = 0$ case where $Jt_b = 0.05$. In
Fig. \ref{fig:2D}(b) we show results for $\mathcal{M}^x(t)$ following
a quantum quench in the 2D quantum Ising model with $N = 5 \times 5$
and $N = 11 \times 11$ sites. In the former case, the results are in
excellent agreement with those obtained via QuSpin's ODE Solver
\cite{Weinberg2019} until $Jt_b = 2.56$; this exceeds the $m^z
= 0$ case, which has a breakdown time of $Jt_b = 0.09$. In the
  absence of another method with which to compare, the results for the
  $N = 11 \times 11$ case are compared to those obtained for smaller
  system sizes. The data track each other until the breakdown time
  $t_b$, suggesting that the results for the larger system size are
  reliable. There is a similarly large improvement over the $ m^z = 0$
  case.

\section{Conclusions}
In this work, we have introduced time-evolving Weiss fields into the
stochastic approach for real-time quantum spin dynamics. We have shown that they
can significantly extend the timescales for numerical simulations, in
both one and two dimensions. In the Appendix, we have further
demonstrated that these Weiss fields can be obtained {\em via} the use
of drift gauges in the gauge-P phase space formulation. It would be
interesting to explore this connection in future work. 

\section{Acknowledgements}
We acknowledge helpful discussions with F. Barratt and S. De Nicola. MJB acknowledges stimulating conversations with D. O'Dell and
S. W\"{u}ster at the ICTS (Bengaluru) program on Non-Hermitian Physics
PHHQP XVIII. SEB is supported by the EPSRC CDT in Cross-Disciplinary
Approaches to Non-Equilibrium Systems (CANES) \textit{via} grant
number EP/L015854/1. We are grateful to the UK Materials and Molecular
Modelling Hub for computational resources, which is partially funded
by EPSRC (EP/P020194/1).  The MPO calculations were performed using
the ITensor Library \cite{ITENSOR}.  AGG acknowledges EPSRC grant
EP/P013449/1. MJB acknowledges the support of the London Mathematical
Laboratory. The data for the figures in this work can be obtained at https://doi.org/10.18742/rdm01-765. \\ 

Recently, we became aware of forthcoming work \cite{DeNicolaPrep} which extends \cite{DeNicola2019euclid} to real time.

\newpage
\bibliographystyle{apsrev4-1}


%

\appendix
\section{Fokker--Planck Description} \label{sec:fp}

As highlighted in the main text, we can make contact between the SDEs
employed here and the gauge-P approach
\cite{Drummond1980,Barry2008,Ng2011,Ng2013}, through the use of drift
gauges \cite{Deuar2002}. To see this, we first consider the
Fokker--Planck description of the stochastic approach to quantum spin
systems \cite{Hogan2004}. This will enable us to develop connections to a broad class
of ``phase space'' methods, which describe quantum systems {\em via}
mappings to classical coordinates.

As usual, one may switch between a Langevin-type description of a
stochastic process, and a Fokker--Planck description, by introducing a
probability distribution $P(\xi)$ for the stochastic variables
$\xi$. For a quantum spin system, this can be introduced by means of
the density matrix $\hat{\rho}=\langle\hat{\rho}^s
\rangle_{\phi,\tilde{\phi}}$, where $\hat\rho^s$ is the stochastic
density matrix $\hat \rho^s=|\psi^s(\phi,t)\rangle\langle
\psi^s(\tilde{\phi},t)|$. More explicitly \begin{align} \hat{\rho}^s =
  \hat{U}^s(\phi,t) \ket{\psi(0)}\bra{\psi(0)}
  \hat{U}^{s\dagger}(\tilde{\phi},t),
\end{align} where $\hat{U}^s(\phi,t)$ is the stochastic time-evolution
operator, and we highlight its noise dependence. The density matrix can
also be expressed as an integral over the classical coordinates
$\xi(t)$ and $\tilde\xi(t)$:
\begin{align} \hat{\rho}(t) = \int d^2\xi d^2\tilde{\xi} ~
P(\xi)P(\tilde{\xi}) ~\hat{\rho}^s(\xi,\tilde\xi)
, \label{eq:fpdecomp}\end{align} where $\hat{\rho}^s(\xi,\tilde\xi) =
\hat{U}^s(\xi)
\ket{\psi(0)}\bra{\psi(0)}\hat{U}^{s\dagger}(\tilde{\xi})$, and both
$P(\xi)$ and $P(\tilde\xi)$ satisfy the Fokker--Planck equation
\begin{align} & \frac{ \partial}{\partial t} P(\xi) =
\hat{\mathcal{F}} P(\xi).\end{align} Here, the differential operator
$\hat{\mathcal{F}}$ contains only first and second order derivatives
with respect to the coordinates $\xi(t)$.
Quantum expectation values can be computed within the Fokker--Planck
representation by
\begin{align}
\langle\hat {\mathcal O}(t)\rangle= \int d^2\xi d^2\tilde{\xi} ~
P(\xi)P(\tilde{\xi}) ~ \text{Tr}
\left(\hat{\rho}^s(\xi,\tilde\xi)\,\hat{\mathcal{O}}
\right). \label{eq:fpobs}
\end{align} 
Without loss of generality, we may consider initial states that are
obtained by time-evolution from the spin-down state $\ket{\downarrow}$
\cite{Begg2019,DeNicola2019euclid}: 
\begin{align} \hat{\rho}^s & = e^{-\frac{1}{2}(\chi +\tilde{\chi}^*)} \prod_j \ket{\xi_j^+}\bra{\tilde{\xi}_j^+},\label{eq:pspaceobs} \end{align}
where $ \ket{\xi_j^+} = e^{\xi_j^+ \hat{S}_j^+}\ket{\downarrow}$ is a
spin coherent state  and $\chi\equiv
\sum_j \xi^z_j$.  As we will discuss in Appendices \ref{sec:positiveP}
and \ref{sec:drift}, Eq.~(\ref{eq:pspaceobs}) enables us to
make contact with the representation of $\hat\rho$ in the phase space
literature \cite{Barry2008,Deuar2002}.  The Weiss-field $m_j^a$ is interpreted as a drift gauge parameter, which we discuss in Appendix
\ref{sec:drift}. To make the connection more explicit we use the
parameterization $z_j = \ln(\xi^+_j)$ and $\omega = -\frac{\chi}{2}$,
as used in Refs \cite{Barry2008,Ng2013}. In this representation
\begin{align} \hat{\rho}^s & = e^{\omega +\tilde{\omega}^*} \prod_j\ket{z_j} \bra{\tilde{z}_j},\label{eq:pspaceobs_line2} \end{align}
where \begin{align}\ket{z_j} = e^{e^{z_j}\hat{S}_j^+} \ket{\downarrow} = \ket{\downarrow} + e^{z_j} \ket{\uparrow} , \label{eq:spincoherentstates}\end{align}
and  $z_j\in {\mathbb C}$.

\section{Phase Space Representations} \label{sec:positiveP}

In this section we give a brief introduction to the phase
space methods developed in Refs
\cite{Drummond1980,Deuar2002,Barry2008,Ng2011,Ng2013}. The initial
starting point is to consider a general parameterization of a density
matrix $\hat\rho$ in terms of phase space variables
$\lambda$: \begin{align} \hat{\rho} = \int d{\lambda} ~ W({\lambda})~
  \hat{\Lambda}({\lambda})
\label{eq:decomp}, \end{align} where $W(\lambda)$ is a
quasi-probability distribution and $\hat{\Lambda}(\lambda)$ is an
operator kernel. Since, $W(\lambda)$ can be negative, it cannot be
interpreted as a true probability distribution. However, for bosons
\cite{Drummond1980} and spins \cite{Barry2008}, $W(\lambda)$ can be
made positive by using a generalized kernel built
from off-diagonal coherent state projectors: $\hat{\Lambda}(\lambda,\lambda') = \prod_j \ket{\lambda_j}\bra{\lambda_j'}$, where $\lambda,\lambda'\in {\mathbb
  C}$ \cite{Drummond1980,Barry2008}. For example,
\begin{align}
\hat{\rho} = \int d^2{\lambda}d^2\lambda' ~
P(\lambda,\lambda') ~ \prod_j \frac{\ket{\lambda_j}\bra{\lambda_j'}}{{\braket{\lambda_j'}{\lambda_j}}}
\label{eq:firstrep}, \end{align} 
where $P(\lambda,\lambda^\prime)$ is positive definite. For bosonic systems, the decomposition (\ref{eq:firstrep}) in which the normalization is explicitly enforced is known as the positive-P representation \cite{Drummond1980}. Analogous representations for spin systems have been considered in Refs \cite{Barry2008,Ng2013}. Phase space distributions over coherent states are not unique due to the overcompleteness of the basis. This can be exploited by using a more general
representation that includes a complex weight $\Omega$. This enlarges
the variable space:
\begin{align}
\hat{\rho} = \int d^2{\lambda}d^2\lambda'd^2\Omega ~ P(\lambda,\lambda',\Omega) ~ \Omega \prod_j \ket{\lambda_j}\bra{\lambda_j'}
\label{eq:decomppos}. \end{align} 
With the inclusion of the weight this is referred to as the gauge-P representation \cite{Deuar2002}. 
This mirrors Eq.~(\ref{eq:fpdecomp}) and Eq.~(\ref{eq:pspaceobs_line2}),
provided we identify $\Omega = e^{-\frac{1}{2}(\chi
  +\tilde{\chi}^*)}$. The integrations over $\lambda$ and
$\lambda^\prime$ can be further identified as the forwards and
backwards time-evolutions involving $\xi^+$ and
$\tilde\xi^+$. Physical observables are calculated according to \begin{align} \langle \hat{{\mathcal O}} \rangle
  & = \int d^2{\lambda}d^2\lambda'd^2\Omega ~ P(\lambda,\lambda',\Omega)~
  \Omega ~ \text{Tr}\left(
  \hat{\Lambda}(\lambda,\lambda^\prime)\hat{{\mathcal
      O}}\right), \end{align} in conformity with Eq.~(\ref{eq:fpobs})
in the stochastic approach.

Having introduced a formal representation of the density matrix, one
may obtain the Fokker--Planck equation by substituting
$\hat\rho$ into the Liouville equation
\begin{align} i \dot{\hat{\rho}} = [\hat{H}, \hat{\rho}] \label{eq:liouville},\end{align}
where $\hat H$ is represented by a differential operator acting
on the classical coordinates. 
Assuming that the Hamiltonian contains no derivative terms higher than second order, one obtains \cite{Barry2008}
\begin{align}  \frac{\partial}{\partial t} P(\uplambda)  = \Big[V + \sum_{j} \frac{
\partial}{\partial \uplambda_j}\Big(-A_{j} + \frac{1}{2}\sum_{l}
    \frac{ \partial}{\partial \uplambda_l}
    D_{jl} \label{eq:fokkerplanck} \Big)\Big]
  P(\uplambda), \end{align} where
$\uplambda=\{\lambda,\lambda^\prime\}$ and we neglect boundary terms
in performing partial integrations. For $V=0$ this is a Fokker--Planck
equation, where $A_{j}$ is the drift vector and $D_{jl}$ is the
diffusion matrix.  The mapping to stochastic Langevin equations can be
carried out provided a ``noise matrix'' $B_{j k}$ exists satisfying
$D_{jl} = \sum_k B_{j k} B_{l k}$ \cite{Drummond1980}. The resulting
Langevin equations in Ito form are given by
\cite{Risken1984}\begin{align} \dot{\uplambda}_{j} = A_{j} + \sum_k
  B_{j k} \phi_k,\label{eq:langevin} \end{align} where $\phi_k$ is
Gaussian white noise satisfying
\begin{align}
\langle \phi_k(t)\phi_{k'}(t')  \rangle = \delta_{k k'} \delta(t-t'), ~~~~~~\langle \phi_k(t)  \rangle = 0.
\end{align}
Observables can be calculated as averages over the noise
\begin{align}  \langle \hat{O} \rangle = \Big \langle \Omega  \text{Tr}\left(\hat{\Lambda}\hat{O} \right) \Big\rangle_{\phi} , \label{eq:generalobs} \end{align}
where $\phi$ includes the forwards and backwards time-evolution, and
$\Omega = 1$ is the un-weighted case.

\section{Drift Gauges} \label{sec:drift}

As discussed in Appendix \ref{sec:positiveP}, the phase space distribution is not unique. The use of different gauges, enables one
to move between these representations. The gauges can be introduced by
adding a vanishing term to the Liouville equation
(\ref{eq:liouville}).
Denoting $\Omega = e^{\omega}$, the identity
$(\frac{\partial}{\partial \omega} - 1) e^{\omega} \hat{\Lambda} = 0$
\cite{Deuar2002} allows one to add \begin{align} & \int
  d^2{\uplambda}d^2\omega P(\uplambda,\omega) f(\uplambda,\omega)
  \Big(\frac{\partial}{\partial \omega} - 1\Big)e^{\omega}
  \hat{\Lambda} = 0 ,\label{eq:addgauge}\end{align} where
$f(\uplambda,\omega)$ is an arbitrary function. To produce a valid
Fokker--Planck equation, without introducing additional noises,
$f(\uplambda,\omega)$ can be constrained \cite{Deuar2002}:
\begin{align} & f(\uplambda,\omega) = V(\uplambda) + \frac{1}{2} \sum_k g_{k}(\uplambda)^2  \frac{\partial}{\partial \omega} + \sum_{k \alpha } g_{k}(\uplambda)B_{\alpha k}(\uplambda)\frac{\partial}{\partial \uplambda_\alpha}   \label{eq:gaugefunction}. \end{align} \normalsize
The first term eliminates $V(\lambda)$ from
Eq.~(\ref{eq:fokkerplanck}), yielding an equation with only first and
second derivatives. The functions $g_k(\uplambda)$ are known as
``drift gauges'' \cite{Deuar2002} as they modify the drift terms in
Eq.~(\ref{eq:langevin}). The resulting Fokker--Planck equation is of the
form (\ref{eq:fokkerplanck}), but with $A_{\alpha} \rightarrow
A_{\alpha} - \sum_{k } g_{k}B_{\alpha k}$ and $D_{\alpha \beta}
\rightarrow D_{\alpha \beta}$ left unchanged.
One must also introduce additional drift and diffusion terms for the
weight variable $\omega$:
\begin{align} & A_{\omega} =  V - \frac{1}{2} \sum_k g_{k}^2,\quad 
D_{\omega \omega} = \frac{1}{2} \sum_k g_{k}^2,
\\ &
D_{\alpha \omega} = D_{\omega \alpha}= \sum_k g_k B_{\alpha k}. 
\end{align}
The drift of
the coherent state parameters $\lambda$ has thus been modified,
via diffusion and drift in the weight variable $\omega$. The modified
Langevin equations are given by
\begin{align}  &  \dot{\uplambda}_{\alpha} = A_{\alpha} - \sum_{k } g_{k}B_{\alpha k} +  \sum_k B_{\alpha k} \phi_k,\label{eq:langevinfinal} \\ &
\dot{\omega} = V - \frac{1}{2} \sum_k g_{k}^2 + \sum_k g_{k}
\phi_k.\end{align} In Appendix \ref{sec:spincoherent}, we use this
formalism to link the Weiss field $m_j^a$ to the drift gauge
$g_k(\uplambda)$.

\section{Spin Coherent States} \label{sec:spincoherent}

In order to make the discussion in Appendices \ref{sec:fp}, \ref{sec:positiveP} and \ref{sec:drift} more
explicit, we introduce spin coherent states following Refs
\cite{Barry2008,Ng2013}. We consider the spin-$\frac{1}{2}$ state decomposition 
\begin{align} \ket{\psi} = \int d^2{z} d^2\omega ~P({z},\omega)~  e^{\omega}  \prod_j  \ket{z_j},\label{eq:waveansatz}\end{align}
where $\omega$ is a complex weight and $\ket{z_j}$ are the un-normalized coherent states defined by (\ref{eq:spincoherentstates}). The spin operators are represented
by differential operators acting on the coherent state parameters:
\begin{align}& \hat{S}^z_j  \ket{z,\omega} = \left(-\frac{1}{2} + \frac{\partial}{\partial z_j} \right) \ket{z,\omega}, \label{eq:zcor} \\
& \hat{S}^x_j \ket{z,\omega} = \left(\frac{1}{2}e^{z_j} - \sinh(z_j)
  \frac{\partial}{\partial z_j} \right)
  \ket{z,\omega}, \label{eq:xcor} \\ & \hat{S}^y_j \ket{z,\omega} =
  \left(\frac{i}{2}e^{z_j} -i \cosh(z_j) \frac{\partial}{\partial
    z_j}\right)\ket{z,\omega} \label{eq:ycor}, \end{align} where
$\ket{z,\omega} \equiv e^{\omega} \prod_j \ket{z_j}$ are weighted
basis states. For a given spin Hamiltonian, we may substitute the
decomposition (\ref{eq:waveansatz}) into the Schr\"odinger equation,
$i \partial_t\ket{\psi(t)} = \hat{H} \ket{\psi(t)},$ in order to
derive the corresponding Fokker--Planck equation for $P(z,\omega)$. In
this representation the analog of (\ref{eq:addgauge}) is \begin{align}
  & \int d^2{z}d^2\omega P(z,\omega) f(z,\omega)
  \Big(\frac{\partial}{\partial \omega} - 1\Big) \ket{z,\omega} = 0
  ,\label{eq:zaddgauge}\end{align} where $f(z,\omega)$ is defined by
(\ref{eq:gaugefunction}). To begin with we set the gauges $g_k(z)$ to
zero. However, $V(z)$ must be chosen to remove the zeroth order
terms. Decomposing $\omega$ into on-site contributions $\omega =
\sum_j \omega_j$ yields the SDEs
\begin{align}
& - i\dot{z}_j = 
 \Phi^z_j  -\Gamma \sinh(z_j) - \frac{1}{2}\sum_{l}J_{jl}, \label{eq:zi}\\ &
-i\dot{\omega}_j = - i V_j(z) \equiv
\frac{\Gamma }{2}e^{z_j} + \sum_l \frac{1}{8}J_{jl} ,  \label{eq:omegai}
\end{align}
where $\Phi^{z}_j = -\sum_k i B_{jk}\phi_k$ is given in terms of
  the independent white noises $\phi_k$, and we have set $V =
  \sum_j V_j$. The ``noise matrix'' $B_{jk}$ is defined \textit{via} the
  diffusion matrix $D_{jl} = iJ_{jl} = \sum_k B_{jk}B_{lk}$. We can
make contact with the SDEs (\ref{eq:plus})-(\ref{eq:zequat}) discussed
in the main text, including the Weiss field $m_j^z$, by introducing
the drift gauge \begin{align} g_k = -\sum_j\left(\frac{1}{2} +
  m_j^z\right)B_{jk}.\end{align} The SDEs (\ref{eq:zi}) and
(\ref{eq:omegai}) become
\begin{align}
& - i\dot{z}_j = 
 \Phi^z_j  -\Gamma \sinh(z_j) , \label{eq:zi2}\\ &
-i\dot{\omega}_j =   
\frac{\Gamma }{2}e^{z_j}  - \frac{1}{2}\Phi_j^z +  i m^{z}_j \sum_k B_{jk} \phi_k,\label{eq:omegai2}
\end{align}	
where now $\Phi^{z}_j = \sum_l J_{jl} m_l^{z} -\sum_k i B_{jk}
\phi_k$. The variables used here can be related to those in the SDEs
(\ref{eq:plus}) and (\ref{eq:zequat}) {\em via} the identification
$z_j = \ln \xi^+_j$, $\omega_j = -\frac{\xi^z_j}{2},$ and $B_{jk} =
\frac{1}{\sqrt{i}} O_{jk}^{zz}$. In writing (\ref{eq:omegai2}) we have
neglected the contribution $\frac{1}{2}\sum_k g_k^2$ since it results
in a deterministic phase for $\ket{\psi^s(t)}$ which is identical for all trajectories.  The additional noise term in
(\ref{eq:omegai2}) appears {\em via} the noise action
(\ref{eq:transformedmeasure}), rather than the SDE (\ref{eq:zequat})
for $\xi^z_j$; the two are equivalent since $\omega$ enters via
$e^\omega$ in (\ref{eq:waveansatz}).

\section{Weiss-Field Calculation} \label{sec:weiss_est}
To calculate the time-evolving Weiss field for the simulations of the
Ising model presented in the main text we use the following
procedure. In the first step we take $\mathcal{N}$ samples of the SDEs
(\ref{eq:isingplus}) and (\ref{eq:isingzequat}) with $m^z_j(t) = 0$,
where we use a two-patch approach to avoid coordinate singularities in
(\ref{eq:isingplus}) \cite{Begg2019}. In the second step, we use these
trajectories to calculate $m^z_j(t)$ \textit{via}
(\ref{eq:hstochwithgauge}). For translationally invariant systems we
consider a single Weiss field obtained from the spatial average
$m^z(t) = \frac{1}{N}\sum_j m_j^z(t)$. We then use this as the local
Weiss field in the next simulation. We repeat steps one and two until
$m^z(t)$ converges to a sufficient level of accuracy over the
time-scales of interest, or until it decays to zero. The resulting
Weiss field can now be used in simulations to obtain quantum
observables. As discussed above, throughout this work we take the
spatial average over the entire system to estimate $m^z(t)$.  However,
as noted in the main text, it can in principle be calculated from a
small sub-system which captures the local interactions between the
spins. To see this, it is convenient to introduce a variant of the
Hubbard--Stratonovich transformation which places decoupling fields on
the bonds between the spins, rather than on the sites of the
lattice.

We consider again the generic quadratic spin Hamiltonian (\ref{eq:heisenbergham}). The interactions in the time-evolution
operator can be decoupled by performing an integral transformation
over auxilliary fields $\eta_{ij}^{ab}\in {\mathbb C}$, which
correspond to the interactions between spins: \begin{align}\hat
  U(t_f,t_i)={\mathbb T}\int {\mathcal D}\eta{\mathcal D}\eta^\ast\,
  e^{-S[\eta,\eta^\ast]+i\int dt \sum_{ja} \Phi_j^a\hat
    S_j^a}.\end{align} Here, ${\mathcal D}\eta{\mathcal
  D}\eta^\ast=\prod^\prime {\mathcal D}\eta_{ij}^{ab}{\mathcal
  D}\eta_{ij}^{ab*}$, where the prime indicates that the product
is over the bonds linking the spins. We label every spin in the array, in arbitrary dimension, with numbers $1$ to $N$, so that $\eta^{ab}_{ij}$ is associated with the bond between sites $i$ and $j$. The effective
magnetic field $\Phi_i^a$ is given by
\begin{align} \Phi_i^a(t) =   \frac{1}{\sqrt{i}} \! \! \sum_{\substack{b j \\ (j<i) }}^{\prime} \! \! \eta^{ab}_{ij} + \frac{1}{\sqrt{i}} \! \! \sum_{ \substack{b j \\ (j>i)}}^{\prime} \! \! \eta^{ba*}_{ji}  + h_i^a,
  \label{eq:hstochabond}\end{align}
where the prime indicates that the summation is restricted to bonds.
The path integral weight is given
  by \begin{align}S[\eta,\eta^\ast]=\int_{t_1}^{t_2} dt
    \sum_{ijab}^\prime \frac{1}{J_{ij}^{ab}} \eta^{ab}_{ij}(t)
    \eta^{ab*}_{ij}(t).\end{align}
The Gaussian ``bond noises'' satisfy
$\langle \eta_{ij}^{ab}(t) \eta_{kl}^{cd*}(t') \rangle = J_{ij}^{ab}
\delta_{ik}\delta_{jl}\delta^{ac}\delta^{bd}\delta(t-t'),$ with
$\langle \eta_{ij}^{ab}(t) \rangle = 0$ and $\langle \eta_{ij}^{ac}(t)
\eta_{kl}^{bd}(t') \rangle = 0$. 
Since the number of bond noises scales with the coordination number,
and each complex noise is the sum of two real noises, they are
computationally more intensive to draw numerically than site-based
noises.  However, bond noises can offer some advantages for Weiss
field estimation. For example, in the case of nearest neighbor
interactions, the spins that are not nearest neighbors will evolve
independently.  In addition, since every spin experiences fluctuations
of the same strength they will have identical mean-field
dynamics if the initial state is translationally invariant. As a
result, in this case it is possible to calculate the time-evolving
Weiss field from a single spin, coupled to its nearest
neighbors. Adding additional spins does not change the stochastic
evolution of the selected spin.

A similar approach to Weiss field estimation can be taken using
site-noise, but it requires additional justification. Recall that the
stochastic magnetic field in direction $a$ experienced by each spin is
given by $\frac{1}{\sqrt{i}} \varphi_j^a$, where $\varphi_j^a =
\sum_{kb} O^{ab}_{jk} \phi_k^{b}$. In general, the matrix
$O_{jk}^{ab}$ ensures that the spins do not evolve independently,
since they experience common noise fields $\phi_k^b$.  However,
only the strength of the noise is relevant for Weiss field estimation
as (\ref{eq:meanfieldit}) is an average for a single spin; the correlations between spins on individual stochastic
trajectories are not required. In particular, if $\sum_{kb} |O_{jk}^{ab}|^2$ is
translationally invariant (i.e. independent of $j$) and we consider
translationally invariant initial states, the Weiss field can be
estimated from $N^{-1}\sum_j m_j^a(t)$. This is true for all the
simulations considered in ths work.

\begin{figure}[t]
\centering
\includegraphics[width = 9.0cm]{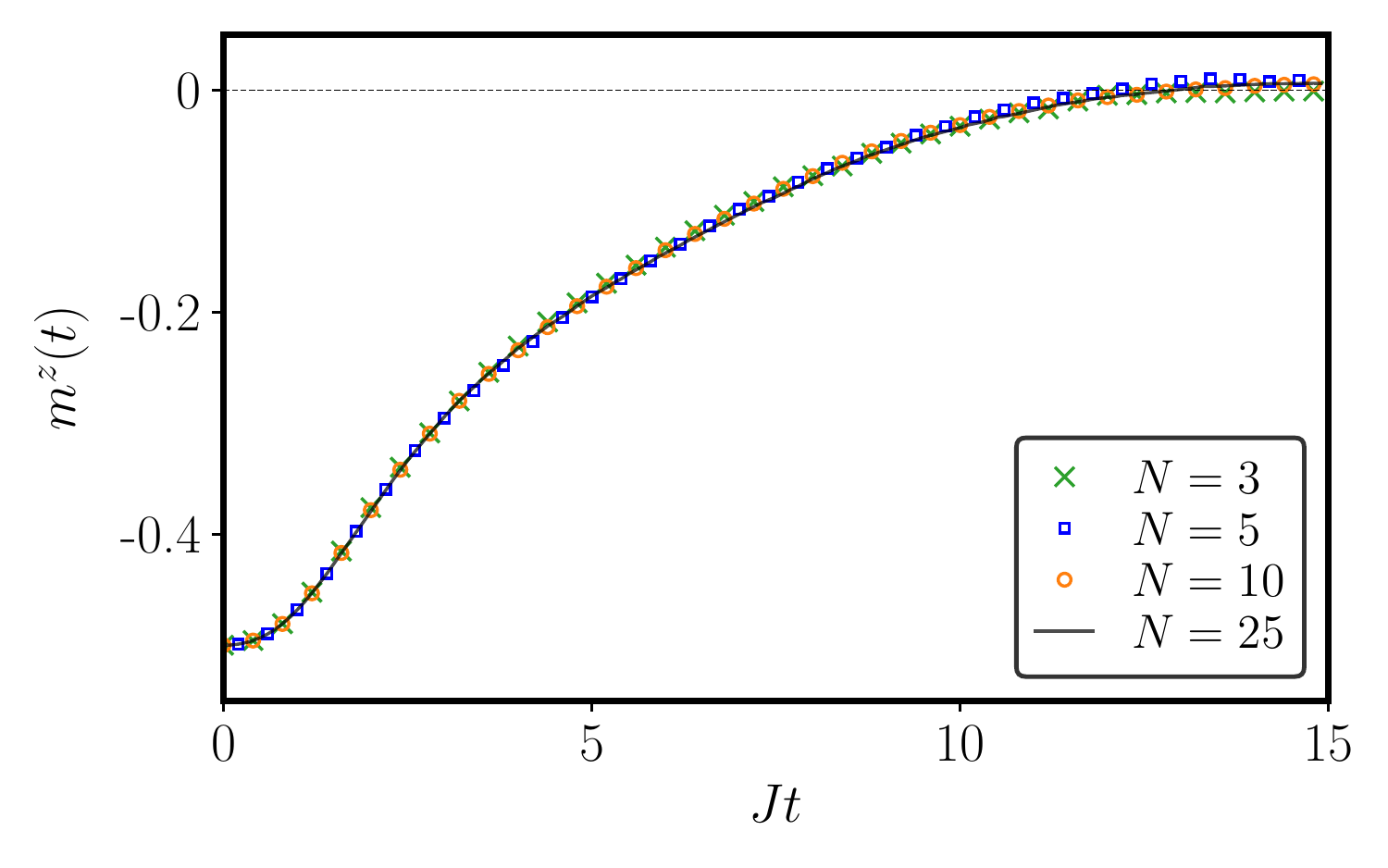}
\caption{Time-dependent Weiss field $m^z(t)$ following a quantum quench in the
  1D quantum Ising model from the fully-polarized state
  $\ket{\Downarrow}$ to $\Gamma/J = 0.3$ for different system sizes.
  The simulations are carried out using site-noise, with four
  iterations of $\mathcal{N} = 5000$ stochastic samples and $dt = 0.1$. Only the
  final iteration is shown.}
\label{fig:Nscaling}
\end{figure}

As in the case of bond-noise, it is possible to estimate the Weiss
field from an appropriate sub-system that reflects the local
interactions, provided that $\sum_{kb} |O_{jk}^{ab}|^2$ is
approximately independent of the system size. For nearest neighbor
interactions, we find that it indeed exhibits only very weak $N$
dependence.  For example, for the $N=101$ site system simulated in
Fig.~\ref{fig:2D}(a), $\sqrt{\sum_k |O_{jk}^{zz}|^2} \approx 1.128$,
while for an $N=3$ spin system $\sqrt{\sum_k |O_{jk}^{zz}|^2} \approx
1.155$, corresponding to a $2\%$ difference. In
Fig.~\ref{fig:Nscaling} we demonstrate the similarity in the extracted
Weiss field for a range of system sizes following a quantum quench in
the 1D quantum Ising model. For the nearest-neighbor 2D simulation of
an $11\times11$ lattice in Fig. \ref{fig:2D}(b) the difference in $\sqrt{\sum_k |O_{jk}^{zz}|^2}$  compared with a $3 \times 3$ system is
approximately $4.3\%$.  Given the effectiveness of a well chosen
static Weiss field, this difference is not expected to be
significant. It should therefore be possible to estimate the Weiss
field by using a smaller system size.

\clearpage

\end{document}